\documentclass[smallextended, prd, aps, color]{svjour3}
\usepackage{graphicx,epsf, epsfig, amssymb}
\usepackage[usenames]{color}

\journalname{GRG}

\newcommand{\be}{\begin{equation}}
\newcommand{\ee}{\end{equation}}
\newcommand{\bel}[1]{\begin{equation}\label{#1}}
\newcommand{\ba}{\begin{eqnarray}}
\newcommand{\ea}{\end{eqnarray}}
\newcommand{\bal}[1]{\begin{eqnarray}\label{#1}}

\def\spose#1{\hbox to 0pt{#1\hss}}
\def\lta{\mathrel{\spose{\lower 3pt\hbox{$\mathchar"218$}}
           \raise 2.0pt\hbox{$\mathchar"13C$}}}
\def\gta{\mathrel{\spose{\lower 3pt\hbox{$\mathchar"218$}}
           \raise 2.0pt\hbox{$\mathchar"13E$}}}

\newcommand{\mbh}{\ensuremath{M_\mathrm{BH}}}
\newcommand{\msigma}{\ensuremath{\mbh-\sigma}}

\begin{document}

\title{Exploring intermediate and massive black-hole binaries with the Einstein Telescope}
\author{Jonathan R.~Gair \and Ilya Mandel \and M.~Coleman Miller \and Marta Volonteri}\institute{J.~R.~Gair \at Institute of Astronomy, University of Cambridge, Cambridge, CB30HA, UK \and I.~Mandel \at Department of Physics and Astronomy, Northwestern University, Evanston, IL, 60208 USA \and M.~C.~Miller \at University of Maryland, Department of Astronomy and Center for Theory and Computation, College Park, MD 20742 \and M.~Volonteri \at Department of Astronomy, University of Michigan, Ann Arbor, MI 48109}

\date{July 26, 2009}

\begin{abstract}

We discuss the capability of a third-generation ground-based detector such as the Einstein Telescope (ET) to enhance our astrophysical knowledge through detections of gravitational waves emitted by binaries including intermediate-mass and massive black holes. The design target for such instruments calls for improved sensitivity at low frequencies, specifically in the $\sim1$--$10$Hz range. This will allow the detection of gravitational waves generated in binary systems containing black holes of intermediate mass, $\sim 100$--$10000M_{\odot}$.  We primarily discuss two different source types --- mergers between two intermediate mass black holes (IMBHs) of comparable mass, and intermediate-mass-ratio inspirals (IMRIs) of smaller compact objects with mass $\sim1$--$10M_{\odot}$
into IMBHs.  IMBHs may form via two channels: (i) in dark matter halos at high
redshift through direct collapse or the collapse of very massive metal-poor Population III stars, or (ii) via runaway stellar collisions in globular clusters.  In this paper, we will discuss both formation channels, and both classes of merger in each case.  We review existing rate estimates where these exist in the literature, and provide some new calculations for the approximate numbers of events that will be seen by a detector like the Einstein Telescope. These results indicate that the ET may see a few to a few thousand comparable-mass IMBH mergers and as many as several hundred IMRI events per year. These observations will significantly enhance our understanding of galactic black-hole growth, of the existence and properties of IMBHs and of the astrophysics of globular clusters. We finish our review with a discussion of some more speculative sources of gravitational waves for the ET, including hypermassive white dwarfs and eccentric stellar-mass compact-object binaries.

\end{abstract}

\maketitle

\section{Introduction} 

The Einstein Telescope (ET), a proposed third-generation ground-based gra\-vi\-ta\-tio\-nal-wave (GW) detector discussed in greater detail elsewhere in this volume, will be able to probe GWs in a frequency range reaching down to $\sim 1$ Hz \cite{Hild:2008,Freise:2009}.  This is lower than the limit of $\sim$40~Hz available to current ground-based interferometric GW detectors such as LIGO, Virgo, and GEO-600 or the $\sim 10$ Hz limit that could be reached by their second generation \cite{LIGO,Virgo,GEO600}.  On the other hand, GWs in the range above $\sim 0.1$ Hz will not be accessible to the planned Laser Interferometer Space Antenna (LISA, \cite{Bender:1998}), which will have sensitivity in the $\sim0.1$mHz--$0.1$Hz range.  The frequency range determines the typical masses of coalescing binaries that could be detected by an interferometer; for example, the frequency of GWs emitted from the innermost stable circular orbit of a test particle around a Schwarzschild black hole of mass $M$ is $\approx 4400\ {\rm Hz} (M_\odot/M)$.  The Einstein Telescope will therefore probe sources with total masses of hundreds or a few thousand $M_\odot$ which are  less likely to be detected by LISA or the current ground-based detectors.  This places the ET in a position to make complementary observations to LISA and LIGO/Virgo/GEO-600 and to carry out unique searches for several very exciting source types, particularly those involving light seeds of massive black holes and intermediate-mass black holes. 

There is a significant body of evidence that massive black holes (MBHs) are generically found in the centers of massive galaxies~\cite{Ferrarese2005}.  These MBHs merge during mergers of their host galaxies, and such mergers therefore trace the history of structure formation in the universe.  Gravitational waves emitted during the mergers of MBHs with masses in the $\sim5\times10^4\ M_\odot$--$5\times10^7M_{\odot}$ will be detectable by LISA;  dozens of detections could be made during the LISA mission \cite{Sesana:2004}.  According to some predictions, these massive black holes grow from light seeds of $\sim 100\ M_\odot$ through accretion and mergers \cite{Madau2001,VHM,SVH:2007}.  The typical frequencies of gravitational radiation emitted during the mergers of such systems will fall in the $0.1$ -- $10$ Hz range, however, and will only be accessible to GW detectors sensitive in that range.  The Einstein Telescope may be able to detect tens of such sources, determining their masses to an accuracy of a few percent and the luminosity distances to $\lesssim 30\%$ \cite{Sesana:2009ET,Gair:2009ET}.

Additionally, globular clusters may host intermediate-mass black holes (IMBHs) with masses in the $\sim 100$ -- $1000\ M_\odot$ range (see \cite{MillerColbert:2004,Miller:2009} for reviews).  Intermediate-mass-ratio inspirals (IMRIs) of neutron stars or stellar-mass black holes into these IMBHs could be detected by the second generation of ground-based detectors \cite{Mandel:2007rates}; however, the Einstein Telescope should be able to detect far greater numbers of events, up to as many as several hundred per year, and these events will range to higher IMBH masses.  If the binary fraction in a globular cluster is sufficiently high or two globular clusters hosting IMBHs merge, an IMBH-IMBH binary can form and then coalesce, emitting gravitational waves~\cite{Fregeau:2006}.  The Einstein Telescope could detect as many as thousands of such events, although, given the present uncertainty about the very existence of IMBHs, all such estimates must be viewed with a great deal of caution.

The Einstein Telescope may also be able to detect a number of other, more speculative, sources.  These include the inspirals of stellar-mass black holes into IMBHs that may reside at the centers of dwarf galaxies, although we do not expect a significant rate of detectable signals of this type.  There is also the possibility that ET will detect orbiting white dwarfs near the upper end of their allowed mass range, or eccentric compact object binaries.

This paper is organized as follows.  In Section \ref{method}, we discuss the methodology for the event-rate calculations.  We describe the adopted detector and network models, the formalism for estimating the signal-to-noise ratio and the waveform families used in the analysis. We then discuss in detail several types of GW sources of particular relevance to the ET. In Section \ref{IMBH}, we consider sources involving IMBHs in globular clusters, both intermediate-mass-ratio inspirals and IMBH-IMBH coalescences. In Section \ref{MBH}, we focus on light massive black holes. We discuss how mergers between galaxies at high redshift can lead to IMBH-IMBH binaries detectable by ET, provided the black hole seeds in the galaxies are light. We also describe how light massive black holes in the centres of dwarf galaxies could act as IMRI sources. In Section \ref{spec}, we discuss several speculative sources, including hypermassive white dwarfs and eccentric binaries.  We finish with a discussion, in Section~\ref{science}, of some of the potential scientific implications of ET observations of these sources. Section~\ref{summary} provides a brief summary.

%\section{Methodology for event-rate and parameter-estimation calculations \label{method}}
\section{Methodology for event-rate calculations \label{method}}

\subsection{The Einstein Telescope configuration\label{ETconfig}}

\begin{figure}[t]
\begin{center}
\includegraphics[width=0.75\textwidth, keepaspectratio=true]{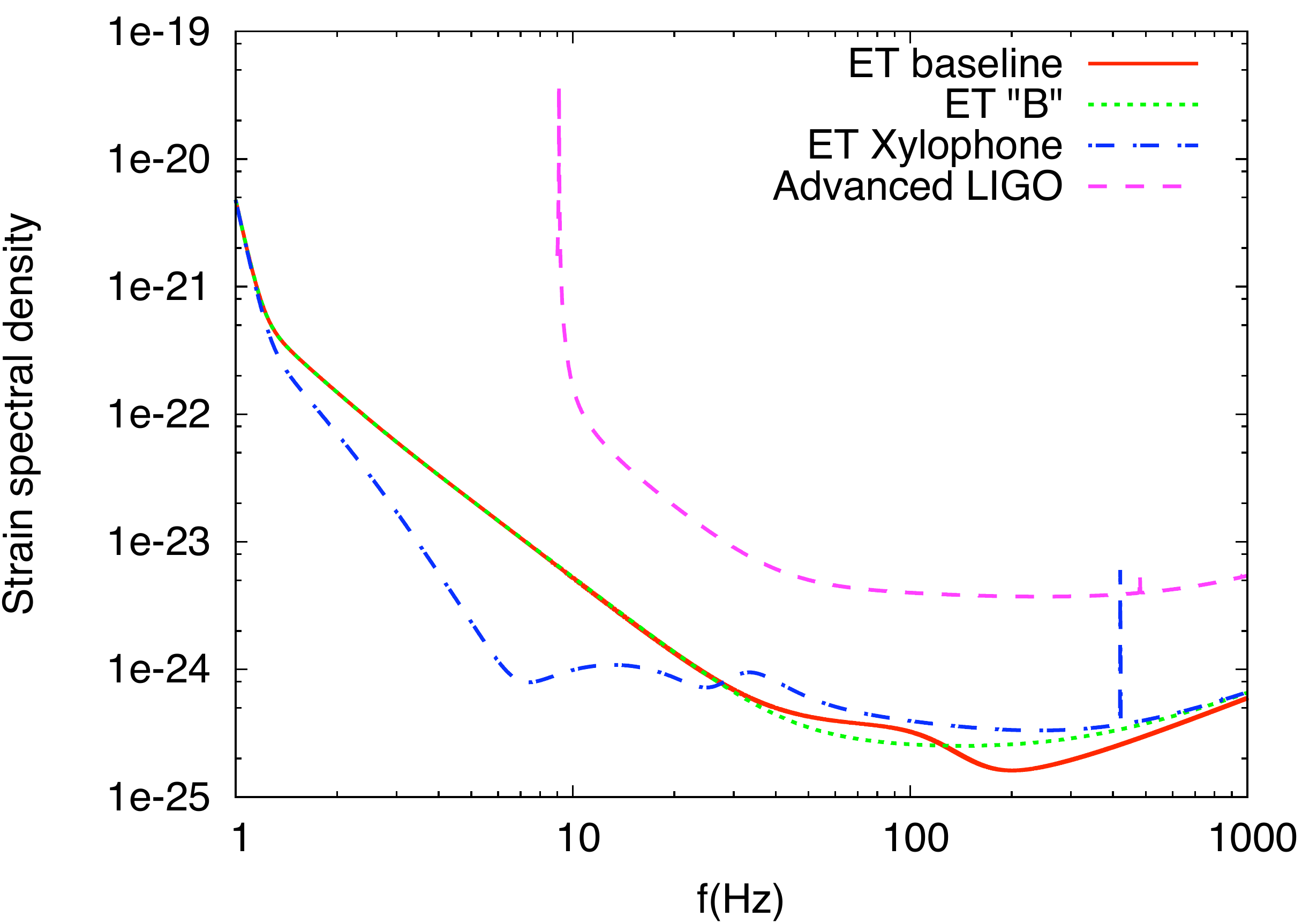}
\end{center}
\caption{Sensitivity curve, $\sqrt{S_h(f)}$, for three configurations of the Einstein Telescope, as described in the text. We also show the Advanced LIGO noise curve for reference.}
\label{ETsh}
\end{figure}

The design target for the Einstein telescope is a 10km scale interferometer, with a factor of $\sim10$ increase in sensitivity over Advanced LIGO, and improved sensitivity at low frequencies. The ET design also calls for the ability to measure polarisation at a single site, which requires at least two non-coaligned coplanar detectors at the site. The currently favoured configuration is a triangular facility, with 10km long arms, and containing three independent detectors with 60$^{\circ}$ opening angles, as this has lower infrastructure costs and slightly better sensitivity than two right angle detectors placed at 45$^{\circ}$ to one another~\cite{Freise:2009}. We refer to this triangular design as a ``single ET''. In Figure~\ref{ETsh} we show the target ET noise curve, labelled ``ET baseline''~\cite{Hild:2008}. This noise curve is for a single right-angle interferometer with the ET design sensitivity. Unless otherwise stated, signal-to-noise ratios (SNRs) etc. will be quoted for this configuration. The sensitivities of one $60^{\circ}$ interferometer, two right-angle interferometers and a single ET are changed relative to this by factors of $\sqrt{3}/2$, $\sqrt{2}$ and $3/2$ respectively. The ``ET baseline'' design has recently been superseded by the curve labelled ``ET B'' in Figure~\ref{ETsh}, but we have checked that this change does not significantly affect our results, since the noise curves are largely similar in the frequency range where massive systems accumulate most of their SNR. Figure~\ref{ETsh} also shows an alternative `xylophone' configuration for ET that was described in~\cite{Hild:2009}. Such a noise curve is realised by operating two detectors within the same vacuum system, one optimised for low-frequency sensitivity and the second for high frequency sensitivity. The net effect on the composite noise curve is an improved sensitivity near $10$Hz. The ET design and noise curve has not yet been finalised, nor what fraction of time ET would spend in high-frequency and low-frequency operation under the xylophone configuration. We will see in Figure~\ref{ETseedratefig} in Section~\ref{seeds} that the xylophone mode is to be preferred for the detection of black hole binaries in the $100$--$1000M_{\odot}$ range.

In Section~\ref{seeds} we will quote results for the parameter estimation accuracies that are achievable for mergers of light seeds of massive black holes detected by ET. As the events are short lived, parameter estimation requires the existence of a network containing multiple detectors. The results we will present are taken from~\cite{Sesana:2009ET,Gair:2009ET}, in which four third-generation network configurations were considered --- (i) one ET at the geographic location of Virgo, plus a second right-angle 10km detector at the location of LIGO Hanford or Perth (Australia); (ii) as configuration (i) plus a third 10km detector at the location of LIGO Livingston; (iii) as configuration (i) but with the Hanford/Perth 10km detector replaced by a second ET; and (iv) three ETs, one at each of the three sites.

%\subsection{Signal-to-noise ratio and parameter estimation \label{SNR}}
\subsection{Signal-to-noise ratio\label{SNR}}

In Sections~\ref{GCimri} and~\ref{GCimbh} we will quote signal-to-noise ratios for events detected with ET. In this section, we describe how these were calculated. The signal-to-noise ratio (SNR) $\rho$ for a waveform $h(t)$ measured by a single detector with one-sided noise power spectral density $S_n(|f|)$ is given by $\rho^2 = \langle h | h \rangle$. Here 
$\langle a | b \rangle$ is the noise-weighted inner product 
\bel{eq:SNR}
\langle a | b \rangle \equiv 4 \Re \int_0^\infty \frac{\tilde{a}(f) \tilde{b}(f)^*}{S_n(|f|)} df,
\ee
where $\tilde{a}(f)$ is the Fourier transform of the waveform $a(t)$ and $^*$ denotes the complex conjugate.

We define the horizon distance $D_{\rm hor}$ as the distance at which an optimally oriented, overhead source produces an SNR of $8$.  The actual gravitational-wave emission is not isotropic and nor is the response of the detector.  We can define the average range as the radius of a sphere whose volume is equal to the true (non-spherical) volume in which inspiral sources can be detected with an SNR greater than $8$.  For uniformly distributed, randomly oriented sources, and without applying corrections for cosmological redshift, this range is related to the horizon distance by $D=D_{\rm hor}/2.26$~\cite{FinnChernoff:1993}. In general, the relationship between $D$ and $D_{\rm hor}$ for sources at cosmological distances will depend on the cosmology, the redshift dependence of the source distribution and even on the intrinsic luminosity of the sources. In this paper, we will ignore these issues and use the usual $1/2.26$ correction factor, since a more rigorous calculation is not currently available. We expect that the error that arises from this approximation will be of the same order as the SNR uncertainties that arise from inaccuracies in the waveform model, which we will discuss in the next section, but emphasise that this must be verified by a proper calculation in the future.  

%To determine the accuracy of parameter estimation, we employ a formalism based on the Fisher information matrix,
%\bel{eq:FIM}
%\Gamma_{ij} = \langle \frac{\partial h}{\partial \theta_i} | \frac{\partial h}{\partial \theta_j} \rangle,
%\ee
%where $\vec{\theta}$ is the vector of model parameters.  However, the Fisher-matrix approach can over-predict the accuracy of parameter estimation when SNRs are low and the parameter space has multiple islands with nearly degenerate waveforms \cite{Vallisneri:2008}.  In addition to these statistical errors, systematic errors in parameter determination can be caused by imperfect knowledge of the waveform used for parameter estimation \cite{CutlerVallisneri:2007}, or by failing to include salient features such as spin in the waveform templates \cite{vanderSluys:2009}.

%For multiple detectors, the inner products in equations (\ref{eq:SNR}) and (\ref{eq:FIM}) should be replaced by sums of inner products over individual detectors. Thus, the existence of $N$ identical interferometers will increase the range by a factor of $\sqrt{N}$ over the single-interferometer range if the network SNR threshold is fixed.

For multiple detectors, the inner product in equation (\ref{eq:SNR}) should be replaced by sums of inner products over individual detectors. Thus, the existence of $N$ identical interferometers will increase the range by a factor of $\sqrt{N}$ over the single-interferometer range if the network SNR threshold is fixed.

\subsection{Waveforms}\label{waveforms}
In this paper, we will focus on two different types of source --- binaries consisting of two intermediate mass black holes of comparable mass; and intermediate-mass-ratio inspirals (IMRIs) of stellar mass compact objects (neutron stars or black holes) into IMBHs. We model the gravitational waves emitted by these systems in different ways. For comparable mass binary systems at the upper end of the detectable mass range, a significant amount of energy is radiated during the merger and ringdown phases and so it is important to include these in waveform models for signal-to-noise ratio calculation and parameter estimation. The recent advances in numerical relativity have allowed the construction of hybrid waveform models that include inspiral, merger and ringdown in a self-consistent way in a single template. There are two models currently available, both of which are for systems containing non-spinning black holes. The assumption of zero spin and the intrinsic waveform uncertainties lead to uncertainties in the SNRs at the level of a few tens of percent, which we will discuss later. However, the corresponding uncertainties in the event rates are much smaller than the typical uncertainties that arise from the astrophysics, since the lower limit on the number of intermediate-mass black holes in the Universe is zero.

The non-spinning ``phenomenological'' inspiral, merger, ringdown model (NSphenom)~\cite{Ajith:2008,Ajith:2008rev}, is constructed by taking a simple ansatz for the waveform model, inspired by post-Newtonian theory, and fitting the coefficients in this model to numerical relativity simulations. The phenomenological model has recently been extended to spinning black holes with non-precessing spins~\cite{Ajith:2009spin}. The ``effective-one-body, numerical relativity'' (EOBNR) model \cite{Buonanno:2007EOBNR} uses post-Newtonian expressions to model the inspiral radiation, which are matched onto fits to numerical relativity simulations for the merger radiation and then onto analytic expressions for the quasinormal mode ringdown radiation. The EOBNR waveforms are constructed to match perturbative results in the test-particle limit where the mass-ratio tends to zero. Denoting the mass of the most massive object in the binary by $m_1$ and the less massive object by $m_2$, the NSphenom and EOBNR models give waveforms as a function of the total mass of the binary, $M=m_1+m_2$, the reduced mass ratio, $\eta = m_1m_2/M^2$ and the time of merger, $t_0$. Including the detector response introduces six additional extrinsic parameters expressing the relative location and orientation of the source and detector: the distance, the two sky-location angles, two binary orientation angles, and the phase at some fiducial time, e.g., $t_0$.

In Section~\ref{seeds} we quote results for the SNRs of comparable mass IMBH mergers detected by ET. These SNRs were computed in~\cite{Sesana:2009ET} using the NSphenom model. ET SNRs for the same systems were recomputed in~\cite{Gair:2009ET} using both models and the results were found to agree to $\sim20\%$, with the NSphenom predictions being higher for low mass ratios, $\eta \approx 0.16$, and the EOBNR predictions being higher for high mass ratios, $\eta \approx 0.25$. The two waveform families are constructed in different ways and have been matched to different numerical relativity simulations, so it is not surprising that the results they predict differ. The level of difference provides a guide to how much the SNRs computed using either waveform family will differ from the true SNRs of these systems that nature provides. This $\sim20\%$ SNR uncertainty from the waveform model must be compared to other uncertainties, such as the omission of spin from the waveform model and uncertainties in the intrinsic astrophysical rates. The effect of spin on the average SNR will also be at the $\sim10\%$ level~\cite{Mandel:2007spin}, although it is more significant for the highest mass systems as it can bring otherwise undetectable sources into the frequency band of the detector. However, the net effect will be to increase the rate of detectable events so using non-spinning models can be thought of as conservative in that regard. A $50\%$ SNR uncertainty leads to an uncertainty in the event rate of a factor of $\sim3$, but typical uncertainties in the astrophysical rates are an order of magnitude. The SNR could be out by as much as a factor of $2$ before it would be comparable to these astrophysical uncertainties, so we consider the waveform uncertainties as fairly negligible. Recent comparisons with additional numerical data suggest that the errors in the EOBNR waveforms are significantly smaller than those in the NSphenom waveforms, at least for equal-mass sources \cite{Buonanno:2009EOBNR}. For this reason, we use the EOBNR model to compute the new results presented in this paper, specifically the SNRs for IMBH-IMBH mergers in globular clusters, which we will describe in Section~\ref{GCimbh}.

The NSphenom and EOBNR waveforms have been matched to numerical relativity simulations, but only for mass ratios of 1:4 and higher. Computational requirements suggest it is unlikely that numerical simulations using current techniques will go beyond mass ratios of $\sim$1:10 in the near future, although this may be possible using innovative new approaches. Post-Newtonian theory also breaks down once the mass-ratio becomes too extreme. For very extreme mass ratios, $\eta \sim 10^{-6}$--$10^{-4}$, gravitational waveforms can be computed using black hole perturbation theory~\cite{Poisson:2004}, in which the smaller object is regarded as a perturbing field of the background spacetime of the larger object and radiation reaction is described in terms of the `self-force'. Significant progress has been made over the past few years in self-force calculations, which has led to the calculation of the self-force for circular orbits in the Schwarzschild spacetime~\cite{BarackSago:2007}, including the shift in the location of the innermost-stable-circular-orbit (ISCO) that results from the action of this force~\cite{BarackSago:2009}. However, even at mass ratios of $\sim10^{-5}$, the terms that are missing in the first-order self-force formalism are estimated to have a marginal effect on the phasing of waveforms for LISA sources~\cite{Huerta:2009}. The mass ratios for typical IMRI sources for ET lie somewhere between these extremes, being typically $\sim 0.001$--$0.1$. In this regime neither post-Newtonian nor perturbative waveforms will be adequate on their own to model the true waveforms~\cite{MandelGair:2009}. More research is needed to devise waveforms that are suitable for bridging this gap. In the meantime we must make do with the available waveforms, with the understanding that these are not completely accurate. 

\begin{figure}[t]
\begin{center}
\includegraphics[width=0.75\textwidth, keepaspectratio=true]{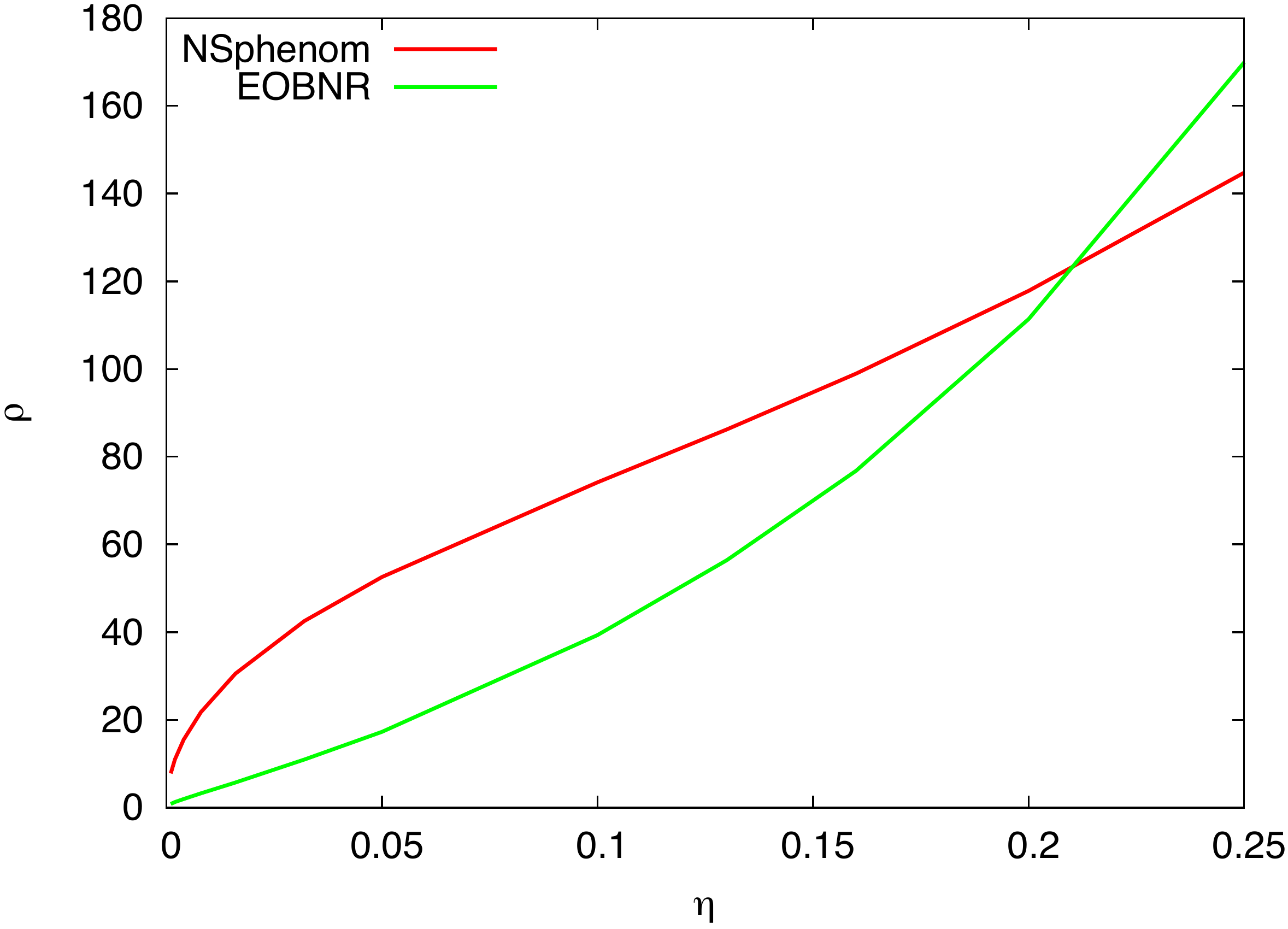}
\end{center}
\caption{Signal-to-noise ratio as a function of mass ratio, $\eta$, computed using the NSphenom and EOBNR waveform families. SNRs are quoted for systems at a luminosity distance of $6.61$Gpc (corresponding to $z=1$) and for a redshifted total mass of $M_z=500M_{\odot}$.}
\label{snrveta}
\end{figure}

In Section~\ref{GCimri}, we will present ET event rates for IMRI sources. These were estimated using SNRs computed with EOBNR waveforms, since these should be accurate both for $\eta \approx 0.25$ and in the limit $\eta \rightarrow 0$. To check the validity of the results, we also computed SNRs using the NSphenom waveforms and using perturbative waveforms for circular and equatorial inspirals, as described and tabulated in~\cite{FinnThorne:2000}, and as used to estimate SNRs for LISA in~\cite{Gair:2009}. The latter waveform model includes only the inspiral phase, and so we compared those results to the inspiral contribution to the EOBNR and NSphenom SNRs. For the early portion of the inspiral, significantly before the last stable orbit, we found that the SNRs obtained with the three waveform families agreed to within $\sim 10\%$; this difference is of a comparable magnitude to the effect of omitting relativistic corrections in post-Newtonian inspirals.  We find, however, that for more massive systems, when merger and ringdown contribute a significant portion of the SNR, estimates from the NSphenom and EOBNR waveforms differ very significantly, with NSphenom waveforms predicting SNRs that are greater by more than an order of magnitude.  In Figure~\ref{snrveta} we show how the SNRs computed using the two models vary as a function of $\eta$ for a fixed luminosity distance and a fixed redshifted total mass of $M_z=500M_{\odot}$. This figure clearly shows the large difference between the two models for small $\eta$. Our theoretical expectation is that the energy emitted during the ringdown should scale as $\eta^2\times\,M$, so that the ringdown SNR should scale as $\eta$ for a fixed total mass. However, the SNR predicted by the ringdown portion of the NSphenom waveform scales roughly as $\sqrt{\eta}$, and is therefore significantly over-predicted for small $\eta$.  For this reason we used EOBNR waveforms, which exhibit the correct scaling, for estimating IMRI SNRs, with the understanding that there is a clear need for more careful and accurate modelling of IMRI radiation in the future. This is essential not least because detection of these systems will almost certainly rely on matched filtering, for which accurate template waveforms are needed.

%Perturbative estimates are also available for the amount of merger/ringdown radiation generated by infall into a black hole, but only for particles on radial infall orbits (see, for example, ~\cite{cardoso03}). There are no self-consistent estimates in the literature for the amount of radiation generated at the end of an extreme-mass-ratio inspiral, since for mass ratios $\eta \lesssim10^{-4}$, the SNR contributed by the ringdown radiation is negligible. Instead, we have compared perturbative inspiral SNRs to SNRs computed using only the inspiral part of the EOBNR waveforms (in the IMR case described by Eq.~(\ref{imrwave}), this would be the part with $f<f_{\rm merg}$). This provides an estimate of the reliability and uncertainty of the SNRs computed in this way, which we assume, without robust justification, carries over to the merger/ringdown radiation as well. The IMRI SNRs should be treated with some caution, but we anticipate this approach will provide reasonably robust estimates of the SNRs with which ET will eventually detect these sources. There is a clear need for more careful and accurate modelling of IMRI radiation in the future, not least because detection of these systems will almost certainly rely on matched filtering, for which accurate templates waveforms are a necessity.

\section{IMBH sources in Globular Clusters \label{IMBH}}

A particularly exciting possibility for GW astronomy is the observations of intermediate-mass black holes (IMBHs) in globular clusters.  Unlike stellar-mass and supermassive black holes, IMBHs have not been identified dynamically, that is, through the observation of one or more orbiting objects and the application of Kepler's law.  Thus, searches for their existence have relied on circumstantial evidence (for reviews and recent results see \cite{MillerColbert:2004,Trenti:2006,Noyola:2008,MaccaroneServillat:2008,Farrell:2009}).  This evidence includes observed fluxes of radiation that, if isotropic and sub-Eddington, would imply IMBH masses above $100~M_\odot$ in many cases; thermal peaks at a few tenths of a keV, which, with normal scalings, would imply masses of tens to thousands of solar masses; and variability features that, if identified with corresponding features observed in systems containing stellar-mass black holes, would also suggest IMBHs weighing in at hundreds of solar masses.  Depending on the threshold one adopts for IMBH likelihood,
candidates exist in one out of every tens to hundreds of galaxies.  Of course, the actual number of IMBHs is likely to be far greater than this, in the same way that, although we have only about 25 strong candidates for stellar-mass black holes in the Galaxy, the total number is likely to be $\sim 10^8$.  The lack of definitive dynamical evidence for IMBHs means, however, that their existence is still inconclusive and alternate explanations for the observations have been proposed (see \cite{Hurley:2007,Colbert:2008} for recent discussions).

One channel leading to IMBH formation is via the runaway collisions of massive stars on time scales too short to allow for stellar evolution, $\lesssim 3$ Myr \cite{PZwart:2004}. Recent simulations of runaway collisions with the inclusion of stellar winds suggest that winds will prevent the growth of IMBHs in all but the most metal-poor environments \cite{Glebbeek:2009}, although these simulations extrapolate wind rates from much less massive stars and the collision rates are so high that the collision products are likely to be extended bags of gas rather than relaxed stars. Alternatively, IMBHs could form through mergers of stellar-mass black holes in dense subclusters at the cores of globular clusters; however, recoil kicks may eject the products of such collisions from the host globulars \cite{OLeary:2006}.  In addition to runaway collisions and repeated stellar-mass BH collisions, other IMBH formation channels include IMBH growth through gas accretion early in the cluster history \cite{PZMcMillanGieles:2010} and direct collapse of Population III stars (these IMBHs could also be seeds for massive black holes, which we discuss in Section~\ref{seeds}).  We refer the reader to \cite{MillerColbert:2004,Miller:2009} for reviews.  

Numerical simulations of globular clusters suggest that IMBHs could merge with numerous lower-mass compact objects (COs) during the lifetime of the cluster \cite{Taniguchi:2000,MillerHamilton:2002a,MillerHamilton:2002b,MouriTaniguchi:2002a,MouriTaniguchi:2002b,Gultekin:2004,Gultekin:2006,OLeary:2006,Mandel:2007rates}, through a combination of the emission of gravitational radiation, binary exchange processes, and secular evolution of hierarchical triple systems.  For IMBH mass $\lesssim 3000 M_\odot$, the GWs generated during the inspiral of a stellar-mass object (black hole or neutron star, since a white dwarf or a main sequence star would be tidally disrupted) into an IMBH are potentially detectable by the Einstein Telescope.  Ringdown radiation could even be detected from more massive IMBHs. 

When the primordial binary fraction in a globular cluster is sufficiently high, $\gtrsim 10\%$, stellar collisions during binary scattering interactions may lead to the production of two IMBHs in a single cluster, according to Monte Carlo simulations carried out by \cite{Gurkan:2006}.  Since observations and numerical calculations suggest that clusters may be born with large binary fractions (e.g., \cite{Ivanova:2005}), the formation of two IMBHs may be generic in sufficiently dense and massive clusters.  If this happens, the two IMBHs will exchange into a common binary which, after shrinking via dynamical friction and dynamical encounters with other stars, will merge through radiation reaction; all of these processes occur on a timescale of $\lesssim 10$ Myr \cite{Fregeau:2006}.  

Additionally, two IMBHs from different globular clusters may merge during the merger of their parent clusters \cite{Amaro:2006imbh}.  N-body simulations suggest that, as the clusters merge, the IMBHs should form a binary with a peak eccentricity of $\sim  0.8$ \cite{Amaro:2006imbh}, although the residual eccentricity would be negligible by the time the frequency reaches $\sim1$Hz and the system enters the ET band.  The binary should merge on a timescale of a few hundred million years  through a combination of dynamical interactions with stars and gravitational-wave emission.  If the probability of forming an IMBH in a cluster is high, and if clusters merge with a probability $\sim 0.1$--$1$ as discussed in \cite{Amaro:2006imbh}, the rates of mergers of IMBHs originating in different host clusters could be competitive with the single cluster channel (see also \cite{pauIMBH}). 

In the following two subsections, we estimate the rates with which the Einstein Telescope could detect gravitational waves from these two globular cluster channels. In Section~\ref{GCimri} we estimate the rates of detectable inspirals of stellar-mass compact-objects into IMBHs (IMRIs) and in Section~\ref{GCimbh} we estimate the rate of detectable coalescences of IMBH-IMBH binaries formed within a single globular cluster.  These subsections contain detailed derivations which are presented here for the first time; readers who are not interested in the details of these calculations will find the results for IMRIs in Table \ref{IMRItable} and the results for single-cluster IMBH-IMBH binary mergers in Equation~(\ref{IMBHIMBHdetectablerate}).

\subsection{Intermediate-mass-ratio inspirals into IMBHs \label{GCimri}}

In an earlier work, a subset of the authors analyzed the possibility of detecting intermediate-mass-ratio inspirals (IMRIs) of compact objects into IMBHs with Advanced LIGO \cite{Mandel:2007rates}.  In that case, it was found that binary tightening via three-body interactions was the dominant channel that led to IMRIs.  The IMBH, as the most massive object in the cluster, readily switches into compact-object (CO) binaries.  Once a sufficiently hard CO-IMBH binary is formed, the binary will be hardened rather than disrupted by three-body interactions with other stars in the cluster.  Eventually, as the interacting stars take away energy from the binary, the binary will tighten to the point where radiation reaction from gravitational-wave emission becomes dominant and drives the binary to coalesce.  For COs that are neutron-stars or black-holes, it is possible to compute the distance to which the gravitational waves can be detected and convert this into an estimated detection rate.

Here, we repeat that calculation for ET sources, with the following two major changes.  First, we take advantage of the recent development of hybrid waveforms that describe all three phases of the coalescence -- inspiral, merger, and ringdown -- to compute the SNR from the full GW signal, rather than just the inspiral portion. We use the EOBNR waveforms~\cite{Buonanno:2007EOBNR} for this calculation. As discussed in Section~\ref{waveforms}, these waveforms have not been tested for mass ratios below $\sim1:4$, but, unlike the NSphenom waveforms, they do appear to behave correctly in the extreme-mass-ratio limit, $\eta \rightarrow 0$. We use these waveforms as there are no better IMRI models available at the present time, but emphasise that there will be some corresponding uncertainty in the results, at the level of tens of percent. Second, because ET has a low frequency cutoff ($\sim 1$ Hz) that is lower than Advanced LIGO ($\sim 10$ Hz), we consider inspirals into $1000 M_\odot$ IMBHs along with inspirals into $100 M_\odot$ IMBHs.  We note, however, that for higher IMBH masses the IMBH could dominate the dynamics in the center of the cluster and a cusp could be formed around the IMBH, possibly increasing the importance of the direct-capture scenario \cite{Hopman:private}; additional discussion of this possibility can be found in Section 2.3 of \cite{Mandel:2007rates}. 

Approximately $(2\pi/22) M/m_*$ close interactions with stars of mass $m_*$ are required to reduce the semimajor axis of a CO-IMBH binary with IMBH mass $M$ by one $e$-folding \cite{Quinlan:1996}.  Stars come within a distance equal to the semimajor axis separation, $a$, of the binary at a rate
\be
\dot{N} \approx n \left[\pi a \frac{2GM}{\sigma^2}\right] \sigma = 
	3 \times 10^{-7} \frac{n}{10^{5.5}\ {\rm pc}^{-3}} \frac{a}{10^{13}\ {\rm cm}} \frac{M}{100 M_\odot} 
	\frac{10\ {\rm km/s}}{\sigma}\ {\rm yr}^{-1},
\label{encrate}
\ee
where the bracketed expression is the gravitationally focused cross-section, $\sigma$ is the velocity dispersion, and $n$ is the number density of stars in a globular cluster, with fiducial values for core-collapsed globulars taken from \cite{PryorMeylan:1993}.  The last $e$-folding time dominates the hardening rate, so the hardening time-scale is
\be
T_{\rm harden} \approx \frac{2\pi}{22}\frac{M}{m_*}\frac{1}{\dot{N}} 
	\approx 2\times 10^8 \frac{10^{5.5}\ {\rm pc}^{-3}}{n} \frac{10^{13}\ {\rm cm}}{a} 
	\frac{\sigma}{10\ {\rm km/s}} \frac{0.5\ M_\odot}{m_*} \ {\rm yr}.
\label{Thard}
\ee 

Meanwhile, the gravitational-wave merger timescale for a binary of semimajor axis $a$, eccentricity $e$, reduced mass approximately equal to the CO mass $\mu\approx m$, and total mass $\approx M$ is \cite{Peters:1964}
\be
T_{\rm GW} \approx 10^{17} \frac{M_\odot^3}{M^2 m} \left( \frac{a}{10^{13}\ {\rm cm}}\right)^4
		(1-e^2)^{7/2} \ {\rm yr} 
	\approx 10^8 \frac{M_\odot}{m} \left(\frac{100\ M_\odot}{M}\right)^2 
		\left( \frac{a}{10^{13}\ {\rm cm}}\right)^4  \ {\rm yr},
\label{Tgravwave}
\ee
where in the last equality we set $e \approx 0.98$ as the eccentricity after the final three-body encounter, following \cite{Gultekin:2006}.  Minimizing the total merger time $T_{\rm merge}=T_{\rm harden}+T_{\rm GW}$ over $a$, while setting $n$, $\sigma$ and $m_*$ to their fiducial values, allows us to compute the CO-IMBH coalescence rate per globular cluster, $1/T_{\rm merge}$. 

To compute the volume within which the Einstein Telescope can detect such IMRIs, we follow the procedure outlined in Section \ref{SNR}.  We use EOBNR waveforms and ignore the spin of the IMBH, which we expect to be small, $S/M^2 \lesssim 0.3$, after a significant number of minor mergers \cite{Mandel:2007spin}.   We compute the range for a ``single ET'' configuration.
% as described in Section \ref{SNR}.
%We compute the horizon distance for a ``single ET'' configuration, and divide it by $2.26$ to obtain the typical range $D$ \cite{FinnChernoff:1993} although this procedure does not correctly average over sky-location and orientation angles when redshift is important.  
The range is a function of the redshifted masses of the IMBH, $M_z=M(1+z)$, and the compact object, $m_z=m(1+z)$.  After computing the range, we convert it into a redshift, $z$, by inverting the following expression for the luminosity distance \cite{Hogg:1999}:
\bel{DLz}
D_L(z) = D_H (1+z) \left\{\int_0^z\frac{dz^\prime}{\left[\Omega_M(1+z^{\prime})^3+
\Omega_\Lambda\right]^{1/2}}\right\}.
\ee
Here, we implicitly assume a flat universe ($\Omega_k=0$) and use $\Omega_M=0.27$,
$\Omega_\Lambda=0.73$, $H_0=72$~km~s$^{-1}$~Mpc$^{-1}$; and $D_H=c/H_0\approx 4170$~Mpc. 
We assume that the typical source is located near the redshift $z$ that corresponds to the search range, and obtain the source-frame masses by dividing the redshifted masses by $1+z$; we use these source-frame masses to compute the merger timescale $T_{\rm merge}$ from Equations~(\ref{Thard}) and (\ref{Tgravwave}).  

We additionally assume that $10\%$ of clusters form an IMBH and are sufficiently dense to be relevant to the rate calculation, and that globular clusters have a fixed comoving space density of $8.4 h^3 {\rm Mpc}^{-3}$ \cite{PZwart:2000}.  For $h=0.72$, this yields a density of $\sim 0.3 {\rm Mpc}^{-3}$ for relevant clusters.  We compute the comoving volume up to redshift $z$ by integrating the following expression for $dV_c/dz$ \cite{Hogg:1999}, with the cosmological parameters defined above:
\bel{Vc}
\frac{dV_c}{dz}=4\pi D_H^3 \left[\Omega_M(1+z)^3+\Omega_\Lambda\right]^{-1/2}
\left\{\int_0^z\frac{dz^\prime}{\left[\Omega_M(1+z^{\prime})^3+
\Omega_\Lambda\right]^{1/2}}\right\}^2.
\ee
The rate of detectable events can then be estimated as $\sim 0.3 (V_c/{\rm Mpc}^3) / [T_{\rm merge}(1+z)]$, where the factor of $(1+z)^{-1}$ is included to convert the coalescence time measured in the source frame to time measured in the observer frame.

\begin{table}[htb]
\begin{tabular}{c@{\quad}c@{\quad\vline\quad}c@{\quad}c@{\quad}c@{\quad}c@{\quad}c@{\quad}c@{\quad}c}
\hline
$M_z/M_\odot$ & $m_z/M_\odot$ & $D$/Gpc & z & $M/M_\odot$ & $m/M_\odot$ &
$T_{\rm merge}$/yr & $V_c$/Mpc$^3$ & Events/yr\\
100 & 10 & 17 & 2.2 & 31 & 3.1 & $4 \times 10^8$ & $7 \times 10^{11}$ & 175\\%300\\
100 & 2 & 6.2 & 1.0 & 51 & 1.0 & $4 \times 10^8$ & $1.3 \times 10^{11}$ & 55\\%70\\
1000 & 10 & 2.2 & 0.4 & 710 & 7.1 & $9 \times 10^7$ & $1.7 \times 10^{10}$ & 40\\
1000 & 2 & 0.7 & 0.15 & 870 & 1.7 & $1 \times 10^8$ & $1\times 10^{9}$ & 2\\
\hline
\end{tabular}
\caption{``Single ET'' average range, corresponding redshift, source-frame masses, merger timescale, comoving volume within range, and detectable event rate for several combinations of plausible redshifted CO and IMBH masses\label{IMRItable}.}
\end{table}

Table \ref{IMRItable} summarizes the rate predictions for four combinations of $M_z$ and $m_z$.  Although the lack of knowledge about IMBHs and their mass distributions makes it impossible to generate firm predictions, and even a lower limit of zero IMRIs is possible, it appears that ET may detect hundreds of compact-object IMRIs into IMBHs over three years of operation.  If ET is operated in the xylophone configuration these rates would increase further.

\subsection{IMBH-IMBH inspirals \label{GCimbh}}

In this section, we wish to estimate the rate at which the single cluster channel generates IMBH-IMBH binaries that are detectable with the Einstein Telescope. To do this, we follow the event rate calculation for LISA and Advanced LIGO described in \cite{Fregeau:2006}.  Once a pair of IMBHs is formed in a single cluster, they sink rapidly to the center where they form a binary and merge via three-body interactions with the stars in the cluster (see \cite{Fregeau:2006,Amaro:2010} for more details).  Therefore, the rate of IMBH binary mergers is just the rate at which pairs of IMBHs form in clusters. The rate of detectable coalescences is
\bel{IMBHIMBHrategeneral}
R \equiv \frac{dN_{\rm event}}{dt_o}=
\int_{M_{\rm tot, min}}^{M_{\rm tot, max}} dM_{\rm tot}
\int_0^1 dq \int_0^{z_{\rm max}(M_{\rm tot},q)} dz 
\frac{d^4 N_{\rm event}} {dM_{\rm tot} dq dt_e dV_c} \frac{dt_e}{dt_o}  \frac{dV_c}{dz}.
\ee
Here $t_o$ is the time measured in our observer's frame and
$t_e$ is the time measured at the redshift $z$ of the merger;
$M_{\rm tot}$ is the total mass of the coalescing IMBH-IMBH binary and $q\leq 1$ is the mass ratio between the IMBHs;
$z_{\rm max}(M_{\rm tot},q)$ is the maximum redshift to which
the ET could detect a merger between two IMBHs of total mass $M_{\rm tot}$ and mass ratio $q$;
$dt_e/dt_o=(1+z)^{-1}$ is the relation between local time and our observed time,
and $dV_c/dz$ is the change of comoving volume with redshift, given by Eq.~(\ref{Vc}).

We make the following assumptions:

\begin{itemize}

\item IMBH pairs form in a fraction $g$ of all globular clusters.

\item We neglect the delay between cluster formation and IMBH coalescence, since it is expected to be no more than a few tens of millions of years~\cite{Fregeau:2006}.

\item When an IMBH pair forms in a cluster, its total mass is a fixed fraction of the cluster mass,
$M_{\rm tot}=2\times 10^{-3}~M_{\rm cl}$. This assumption is based on what is typically seen in simulations \cite{Gurkan:2004}.  As there are no current constraints on the mass ratio, we take it to be uniformly distributed between $0$ and $1$.  We restrict our attention to systems with a total mass in the IMBH range, which we adopt to be between $M_{\rm tot, min}=100 M_\odot$ and $M_{\rm tot, max}=20000 M_\odot$. This means we confine our attention to clusters with masses $5\times10^4 \leq M_{\rm cl}/M_\odot \leq 10^7$ (note that the lower limit is different from that chosen in \cite{Fregeau:2006} since here we set $M_{\rm tot, min}=100 M_\odot$ for IMBH sources).  
Thus,
\be 
\frac{d^4 N_{\rm event}} {dM_{\rm tot} dq dt_e dV_c} = g \frac{d^3 N_{\rm cl}} {dM_{\rm cl} dt_e dV_c} \frac{1}{2\times 10^{-3}}.
\ee

\item The distribution of cluster masses scales as $(dN_{\rm cl}/dM_{\rm cl}) \propto M_{\rm cl}^{-2}$ independently of redshift~\cite{ZhangFall:1999} and
%We confine our attention to clusters with masses ranging from  $M_{\rm cl, min}=5\times10^4 M_\odot$ to $M_{\rm cl, max}=10^7 M_\odot$.
the total mass formed in all clusters in this mass range at a given redshift is a redshift-independent fraction $g_{\rm cl}$ of the total star formation rate per comoving volume:
\begin{equation}
g_{\rm cl} \frac{d^2M_{\rm SF}}{dV_c dt_e} = \int_{M_{\rm cl, min}}^{M_{\rm cl, max}} 
\frac{d^3 N_{\rm cl}} {dM_{\rm cl} dt_e dV_c} M_{\rm cl} dM_{\rm cl},
\end{equation}
which provides the normalization for $dN_{\rm cl}/dM_{\rm cl}$:
\begin{equation}
\frac{d^3 N_{\rm cl}} {dM_{\rm cl} dt_e dV_c}= \frac{g_{\rm cl}}{\ln (M_{\rm cl, max}/M_{\rm cl, min})} \frac{d^2M_{\rm SF}}{dV_c dt_e} \frac{1}{M_{\rm cl}^2}.
\end{equation}

\item The star formation rate as a function of redshift $z$ is
\begin{equation}
\frac{d^2M_{\rm SF}}{dV_c dt_e}=0.17\frac{e^{3.4z}}{e^{3.4z}+22}
\frac{\left[\Omega_M(1+z)^3+\Omega_\Lambda\right]^{1/2}}
{(1+z)^{3/2}}~M_\odot~{\rm yr}^{-1}~{\rm Mpc}^{-3}.
\end{equation}
This is the formula used by \cite{Steidel:1999}, in which the 
star formation rate rises rapidly with increasing $z$ to $z\sim 2$,
after which it remains roughly constant.  As in Section \ref{GCimri}, we assume
a flat universe ($\Omega_k=0$), and use $\Omega_M=0.27$,
$\Omega_\Lambda=0.73$, and $H_0=72$~km~s$^{-1}$~Mpc$^{-1}$.

\end{itemize}

Under these asssumptions, Eq.~(\ref{IMBHIMBHrategeneral}) predicts
the rate of detectable coalescences per year as
\bal{IMBHIMBHrate}
R &=& \frac{2\times 10^{-3} \  g \  g_{\rm cl}} {\ln (M_{\rm tot, max}/M_{\rm tot, min})} 
\int_{M_{\rm tot, min}}^{M_{\rm tot, max}}  \frac{dM_{\rm tot}}{M_{\rm tot}^2} \int_0^1 dq\\
\nonumber 
&&
\int_0^{z_{\rm max}(M_{\rm tot},q)} dz \ 
0.17\frac{e^{3.4z}}{e^{3.4z}+22}
\frac{4\pi D_H^3} {(1+z)^{5/2}}
\times
\left\{\int_0^z\frac{dz^\prime}{\left[\Omega_M(1+z^{\prime})^3+
\Omega_\Lambda\right]^{1/2}}\right\}^2.
\ea
Note that here $M_{\rm tot}$ is measured in solar masses and $D_H$ is measured in Mpc.

Rather than computing $z_{\rm max}(M_{\rm tot},q)$ for all values of $M_{\rm  tot}$ and $q$, we rely on the following fitting formula for the average range $D$ as a function of the redshifted total mass $M_z=M_{\rm tot} (1+z)$, obtained by using EOBNR waveforms to model the coalescence (see Section \ref{SNR}): 
\be
D (M_z) = (A \ {\rm Mpc})   
	\left\{ \begin{array}{ll}
	\left(\frac{M_z}{M_\odot}\right)^{3/5}&\mbox{if } M_z<M_0 \\
	\left(\frac{M_0}{M_\odot}\right)^{11/10} \left(\frac{M_z}{M_\odot}\right)^{-1/2}&\mbox{if }  M_z>M_0
	\end{array} \right. ,
\ee
where $A=500$, $M_0=600 M_\odot$ for $q=1$ and $A=281$, $M_0=450 M_\odot$ for $q=0.25$. 
%We use $\rho=8$ as the SNR threshold for a ``single ET'' configuration.  We average over sky-location angles and orientations as described in Section \ref{SNR}.
%We also approximate averaging over source sky-location angles and orientations by a factor of $1/2.26$,  even though this conversion factor between the horizon distance and the average range is only applicable when the source distribution is isotropic and redshift corrections are unimportant (see Section \ref{SNR}).

Lensing of gravitational wave sources adds some uncertainty to this picture. Individual sources can be magnified or de-magnified by lensing, making them visible at a greater or lesser distance than predicted by the preceding formula. Flux conservation ensures that the expected magnification of a source is $1$, meaning no net change in flux (this argument was first elucidated in~\cite{weinberg76}), which leads us to expect that the total change in the event rate will be small. The magnification distribution that arises from lensing peaks at less than $1$, i.e., a demagnification, and shows an exponential fall-off for large magnifications~\cite{peacock82,wang02}. The peak moves toward greater demagnification for sources at higher redshifts, but this is compensated by a longer tail toward very high magnifications. The amount of volume added to a flux-limited sample by these highly magnified lines of sight compensates for the smaller volume lost by each of the (greater number) of demagnified lines of sight. In~\cite{peacock82}, it was shown that weak lensing did not significantly change the number counts in a flux limited radio sample, but the number of events tended to be increased by the strong lensing tail. The total change in number counts was only a few percent. We can therefore ignore the effect of weak lensing on the number counts of gravitational wave sources, although it will have an impact on the precision to which distances can be measured with gravitational wave observations.

Ignoring lensing, we can compute $z(D_{\rm L})$ by inverting Eq.~(\ref{DLz}).  For a given choice of $M_{\rm tot}$ and $q$, the maximum detectable redshift $z_{\rm max} (M_{\rm tot}, q)$ is then obtained by finding a self-consistent solution of 
\be
z\Big(D_{\rm L}\big(M_{\rm tot} (1+z_{\rm max})\big)\Big)=z_{\rm max}.
\ee

The integrals over $M_{\rm tot}$ and $z$ in Eq.~(\ref{IMBHIMBHrate}) were evaluated for two specific values of $q$.  For $q=1$, the total rate was found to be $R=7.5\times10^4\ g\ g_{\rm cl}$; for $q=0.25$, it was $R=2.7\times10^4\ g\ g_{\rm cl}$.  The range varies smoothly with $q$ and so we can estimate the integral over $q$ to be the average of these two rates. This yields a final estimate of the total rate as
\bel{IMBHIMBHdetectablerate}
R \approx 500 \left(\frac{g}{0.1}\right) \left(\frac{g_{\rm cl}}{0.1}\right) {\rm yr}^{-1},
\ee
where we arbitrarily chose $g=0.1$ and $g_{\rm cl}=0.1$ as the reference values for these unknown parameters.

%2500, 2000 -> 750, 272

\section{Sources in low-mass galaxies \label{MBH}}

\subsection{Light seeds of MBHs at high redshifts \label{seeds}} 

Supermassive black holes (SMBHs) weighing millions to billions of solar masses are nowadays believed to reside in most local galaxies~(\cite{Ferrarese2005} and references therein).   The masses of today's SMBHs exhibit clear correlations with the properties of their host galaxies (luminosity, mass, and stellar velocity dispersion), suggesting there is a single mechanism for assembling SMBHs and forming galaxies. The evidence therefore favours a co-evolution between galaxies and SMBHs.

In the currently favoured cold dark matter cosmology, galaxies today are expected to have been built up, via a series of mergers, from small-mass building blocks that condensed out at early cosmic times. A single big galaxy can be traced back to hundreds of smaller components with individual masses as low as $\sim 10^5$ M$_\odot$. Similarly, we expect the SMBHs found in galaxies today to have grown partially by accretion and partially by mergers following mergers between galaxies (e.g., \cite{VHM,Malbon2007}), so that a single SMBH can be traced back to some number of `seed' black holes at early times \cite{VLN2008}. There are large uncertainties in this picture, however. Did seed black holes form efficiently in small galaxies (with shallow potential wells) at early times, or was their formation delayed until substantial galaxies with deeper potential wells had been formed? This is a key question, as the mass and the occupation number of the seeds ultimately dictates the occupation number of SMBHs in galactic centers.

The formation of SMBHs is far less well understood than that of their light stellar-mass counterparts. The `flow chart' presented by \cite{Rees1978} still stands as a guideline for the possible paths leading to the formation of SMBH seeds in galactic nuclei. One possibility is  that the seeds of SMBHs were the remnants of the first generation of stars, formed out of zero-metallicity gas \cite{Madau2001}. In a cold dark matter universe, structure builds up hierarchically, so the smaller clumps at the earliest cosmic times have shallower potential wells. Stars cannot form until the clumps are sufficiently big to provide a potential well deep enough to pull in gas that can cool radiatively and contract to make a protostar. This requires dark matter clumps -- minihalos -- of $\sim 10^6$ M$_\odot$ at redshifts of $z \sim 20$. The first stars forming in these minihalos develop under very different conditions from present-day stars: there are no heavy elements (so that molecular hydrogen is the only effective coolant), no dust, and no magnetic fields. These conditions mean that these `Population III'  stars  were likely very massive, having characteristic masses of the order of  $\sim 100$ M$_\odot$  (e.g., \cite{Bromm1999,Nakamura2001,Abel:2002,Yoshida2006}).  This prediction relies on the absence of efficient cooling agents in the primordial metal--free gas.  If Population III  stars form with masses 40 M$_\odot < M< $140 M$_\odot$ or  M$ > 260$ M$_\odot$, they are predicted to  collapse and form IMBHs directly with little mass loss \cite{Fryer2001}, i.e., leaving behind seed IMBHs with masses $M_{BH} \sim 10^2-10^3\,M_\odot$. This is a plausible formation mechanism for the seeds upon which supermassive black holes are grown \cite{VHM},  although more massive black holes may have been formed after the epoch  of the first stars in dark-matter halos with virial temperatures of  $\sim 10^4$ K \cite{Bromm2003,Spaans2006,BVR2006} via `direct collapse', as described in the following. 

Direct collapse models  for MBH formation rely on the collapse of supermassive objects formed directly out of dense gas \cite{haehnelt1993,LoebRasio1994,BrommLoeb2003,Koushiappas2004,BVR2006,LN2006}. The physical conditions (density, gas content) in the inner regions of mainly gaseous proto-galaxies make these loci natural candidates, because the very first proto-galaxies were, by definition, metal-free, or at the very least very metal-poor. Enriched halos have a more efficient cooling, which in turn favours fragmentation and star formation over the efficient collection of gas conducive to MBH formation. In a typical galaxy, however, the tidally induced angular momentum would still be enough to provide centrifugal support at a distance $\simeq 20$ pc from the centre, and halt collapse, ultimately leading to the formation of a disk. Additional mechanisms inducing transport of angular momentum are needed to further condense the gas until conditions fostering MBH formation are achieved. An appealing route to efficient angular momentum shedding is by global dynamical instabilities, such as the ``bars-within-bars" mechanism, that relies on global gravitational instability and dynamical infall \cite{Shlosman1989,BVR2006}.  Self-gravitating gas clouds become bar-unstable when the level of rotational support surpasses a certain threshold. A bar can transport angular momentum outward on a dynamical timescale via gravitational and hydrodynamical torques, allowing the radius to shrink.  Provided that the gas is able to cool, this shrinkage leads to even greater instability, on shorter timescales, and the process cascades.  This mechanism is a very attractive candidate for collecting gas in the centres of halos, because it works on a dynamical time and can operate over many decades of radius. It has also been proposed that gas accumulation in the central regions of protogalaxies can be described by local, rather than global, instabilities. During the assembly of a galaxy disc, the disc can become self-gravitating. As soon as the disc becomes massive enough to be marginally stable, it will develop structures that will redistribute angular momentum and mass through the disc, preventing the surface density from becoming too large and the disc from becoming too unstable. To evaluate the stability of the disc, the Toomre stability parameter formalism can be used \cite{Koushiappas2004,LN2006}.   The gas made available in the central compact region can then form a central massive object, for instance via the intermediate stage of a `supermassive' star \cite{hoyle1963,Baumgarte1999}, or a `quasistar', an initially low-mass black hole rapidly accreting within a massive, radiation-pressure-supported envelope, \cite{BVR2006,Begelman2008}. In both cases, the mass function of seeds is predicted to peak at $10^5-10^6 M_\odot$ \cite{VLN2008}.

\begin{figure}[t]
\begin{center}
\includegraphics[width=0.75\textwidth, keepaspectratio=true]{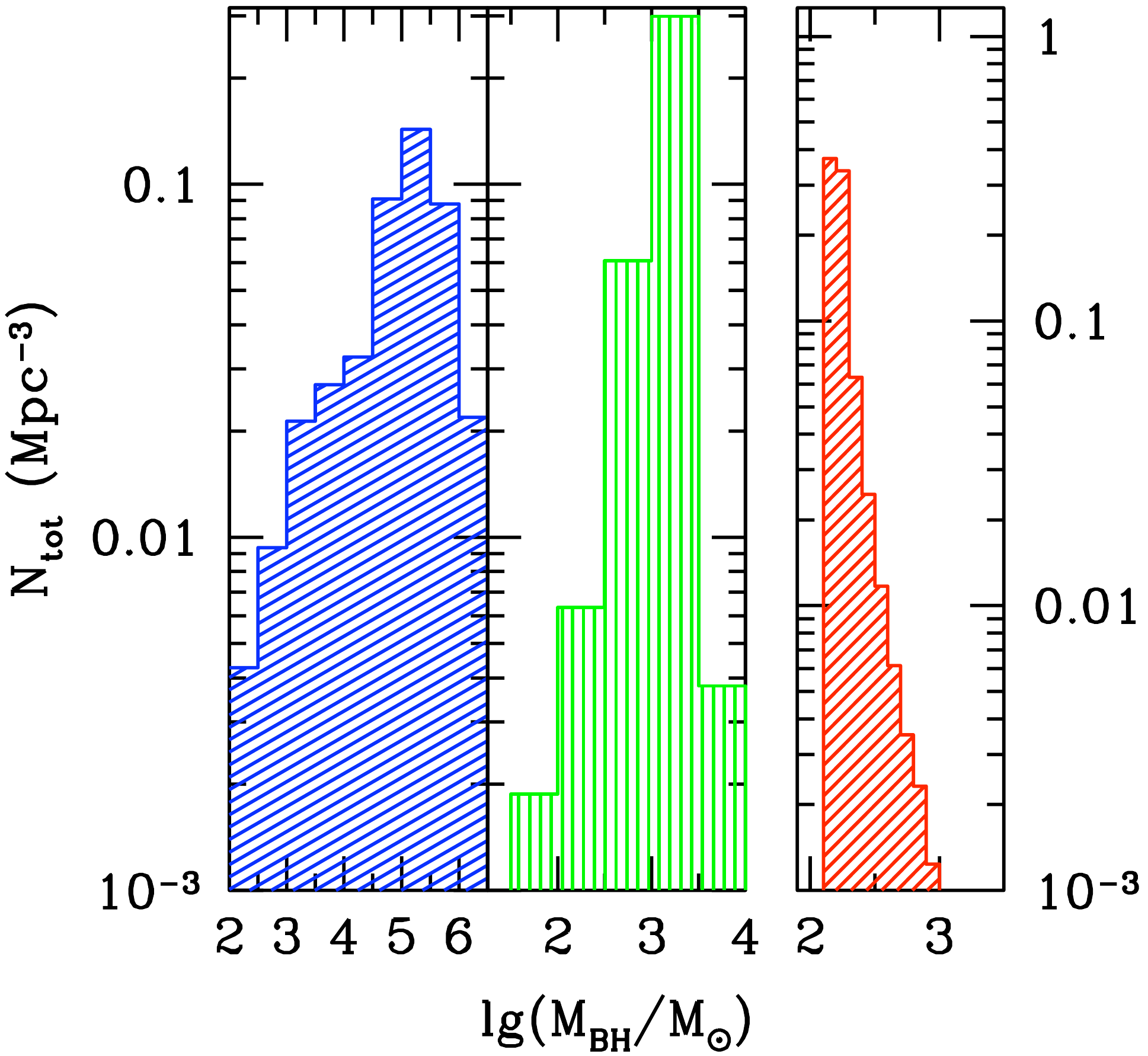}
\end{center}
\caption{Mass function of seed MBHs for three different formation scenarios. Left: direct collapse \cite{VLN2008}; centre: runaway stellar mergers in high-redshift clusters \cite{Devecchi2009}; right: Population III remnants \cite{Madau2001}. Note the different y-axis scale for the Population III case.}
\label{fig:MF}
\end{figure}

As described in Section \ref{GCimri}, the formation of an IMBH as a result of dynamical interactions in dense stellar systems is a long-standing idea, which could also create intermediate mass MBH seeds.  This process could have been very effective in the very first stellar clusters that formed in high-redshift proto-galaxies, when the Universe was not as metal-rich as now.   Low metallicity  favors the growth of a very massive star, the precursor of an IMBH remnant.  The mass loss due to winds is significantly reduced in metal-poor stars, which greatly helps in increasing the mass of the final IMBH remnant (cf.~\cite{yungelson2008}).  The formation of stellar clusters and the possible evolution of the stellar systems up to IMBH formation are explored in \cite{Devecchi2009}. Figure~\ref{fig:MF} shows three mass functions for three different MBH `seed' scenarios: direct collapse \cite{BVR2006}, runaway stellar mergers in high-redshift clusters, and Population III remnants \cite{VHM}.

It is uncertain how many MBH `seeds' formed, and in which mass range. Equally uncertain is how these `seed' black holes grew within their host minihalos.  It is not obvious if efficient accretion onto these seeds could have taken place, at least early on, in the fragile environment that the shallow potential wells of minihalos represent \cite{Milos2009}.  It is likely that seed IMBHs can grow efficiently only if they are hosted in the most massive galaxies at these early cosmic epochs, while IMBHs in an `average' galaxy could have experienced intermittent and inefficient accretion, thus leaving behind a population of underfed IMBHs with a mass range similar to that of the original seeds, $M_{BH} \sim 10^2-10^3\,M_\odot$.

The Einstein Telescope will be able to probe mergers between black-hole seeds at high redshift, and thus help to distinguish between these various channels for seed formation. An estimate for the ET event rate under the Pop III model can be computed using Monte-Carlo merger-tree realizations based on the extended Press-Schechter formalism~\cite{PressSchechter:1974}, as described in~\cite{VHM,Volonteri:2007}. This was done in~\cite{Sesana:2009ET}, and the results which we now quote are all taken from that work and the companion paper~\cite{Gair:2009ET}. These papers considered four different models that were based on the same merger tree realisations (taken from~\cite{VHM,Volonteri:2007}), but differed in the initial mass distribution of seeds and in the prescription for accretion onto the seed black holes. In these scenarios, which were all based on having the light, Pop III, remnants described earlier in this section as the seeds for black hole formation, a single ET would detect $\sim1$--$10$ seed mergers, depending on the model. The detected mergers would be between black holes with total mass ranging from $2M_{\rm min}$ up to $\sim 1000M_{\odot}$, where $M_{\rm min}$ is the mass of the lightest seed black hole in the initial mass distribution. This minimum seed mass is rather uncertain and depends on the details of the model used, as discussed earlier. In the scenarios considered in~\cite{Sesana:2009ET}, $M_{\rm min}$ was either $10M_{\odot}$ or $150M_{\odot}$. The detected events would be seen at redshifts $z\sim1$--$7$, although this could extend to $z\sim12$ for the lightest seed model, which had $M_{\rm min} =10M_{\odot}$. If ET was operated in the xylophone configuration described in Section~\ref{ETconfig}, the number of events seen would be increased to several tens, and these would be out to a redshift $z\sim15$~\cite{Gair:2009ET}.

\begin{figure}[t]
\begin{center}
\includegraphics[width=0.75\textwidth, keepaspectratio=true]{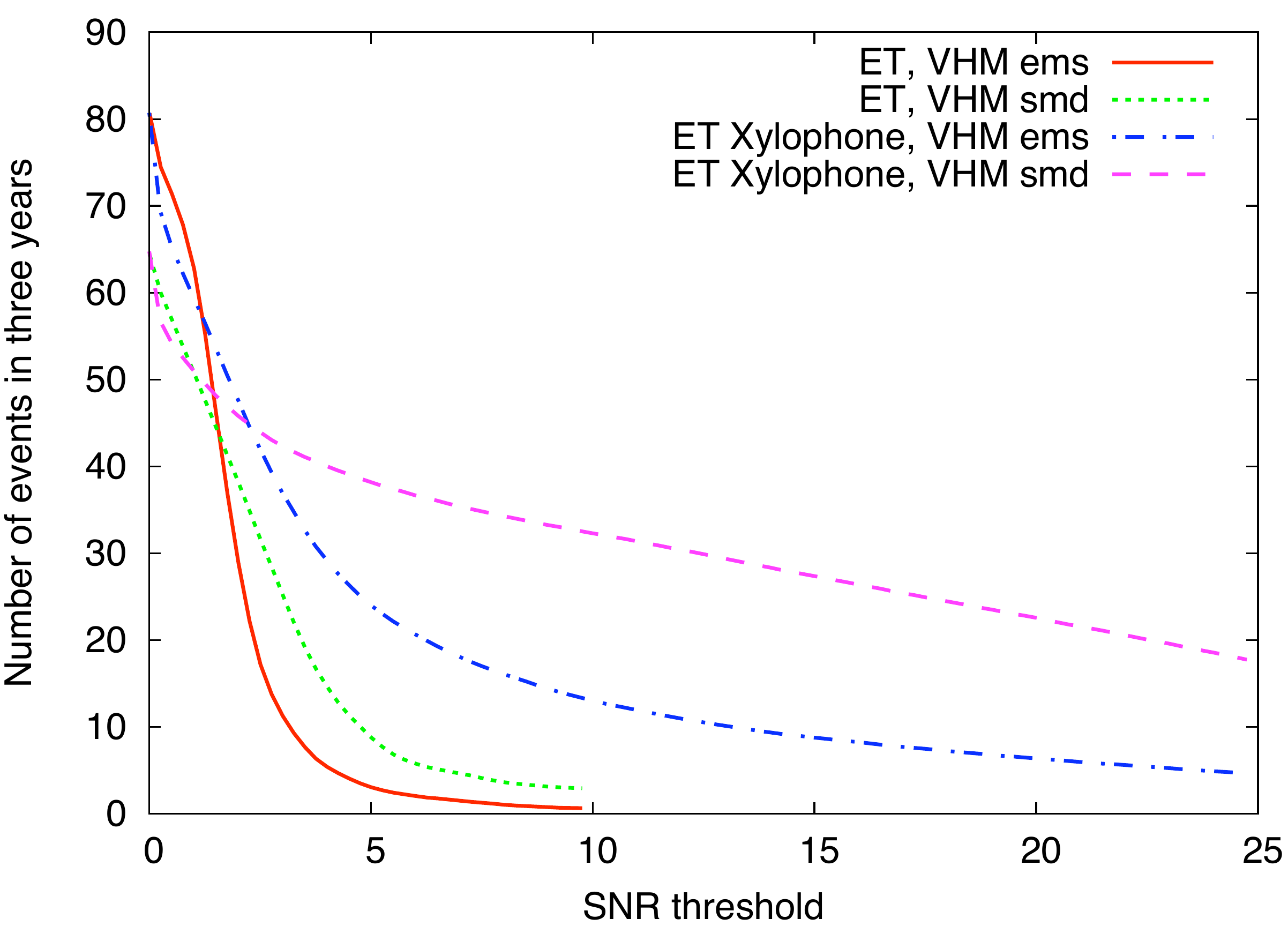}
\end{center}
\caption{Number of events detected by the Einstein Telescope in three years, as a function of the required signal-to-noise ratio threshold in a single right-angle detector. Results are shown for both the baseline and xylophone configurations of the Einstein Telescope, and for two different astrophysical models --- Volonteri-Haardt-Madau (VHM) with equal mass seeds (\emph{VHM,ems}) and VHM with a seed mass distribution (\emph{VHM,smd}). Details on these models can be found in~\cite{Gair:2009ET}.}
\label{ETseedratefig}
\end{figure}

Figure~\ref{ETseedratefig}, reproduced from the data in~\cite{Gair:2009ET}, shows how the number of events seen by ET over three years varies as a function of the signal-to-noise ratio required in a single 10km right-angle interferometer for detection. The SNR required in the network of detectors is likely to be $\sim 8$, although this depends somewhat on data-analysis issues, and on the amount of source confusion in the data stream. A network SNR of $8$ corresponds to an SNR in the single right-angle detector of $5.3$ for a single ET, or SNRs of $4.8$, $3.9$, $3.8$ and $3.1$ for the network configurations (i) -- (iv) described in Section~\ref{ETconfig}. In Figure~\ref{ETseedratefig} we show results for two of the four light-seed models considered in~\cite{Sesana:2009ET}, and for both the baseline and xylophone configurations of the detector. The rate for the baseline ET configuration is rather sensitive to the SNR that is ultimately required for a confident detection, but the xylophone configuration is more robust, as it has improved sensitivity at just the right frequency for systems with mass in the $100$--$1000M_{\odot}$ range. The mergers seen by ET will be complementary to mergers between heavier black holes that will be seen by space-based detectors such as LISA, ALIA or DECIGO~\cite{Gair:2009ET}. The combination of detectors will provide a nearly complete survey of mergers between galactic black holes, yielding important constraints on astrophysical models of galaxy formation and growth. The utility of observations of binary black hole systems with multiple detectors was also discussed in~\cite{pauIMBH}, with specific reference to IMBH-IMBH binary mergers arising from the mergers of globular clusters containing central IMBHs. They found that, if LISA and ET were operating concurrently, the same IMBH binary could be detected by both detectors with a time separation of a few months. Even if observations of the same system are not made with different detectors, each detector will provide a measurement of the rate of black hole mergers in a different black hole mass range. These rate observations over the black hole mass spectrum will provide important constraints on models of black hole growth.

One important question is whether ET will be able to distinguish between black-hole mergers coming from this channel, and those described in Section~\ref{GCimbh} that arise in globular clusters. To provide constraints on merger histories, it is necessary to know that an observed event is associated with a galaxy merger. The masses and redshifts of events will provide some information, but more work is required to understand what observational signatures provide the best discriminating power. We would expect mergers between seed black holes to occur over a range of redshifts, with some events at redshifts $z \gtrsim 10$. In the mechanism described in Section~\ref{GCimbh}, the black-hole binaries form and merge very quickly, so this could also produce events over a range of redshifts. However, the distinction between these two formation channels becomes increasingly vague at high redshift, when galaxies are in the process of formation. What is important for the light-seed scenario is that black holes of low mass, $\sim100M_{\odot}$, exist at high redshift. Therefore, being able to identify an event as being between two $\sim 100M_{\odot}$ black holes at redshift $z\gtrsim 5$ would be an important constraint, regardless of how that seed had initially formed. A single ET cannot measure the six extrinsic parameters of a merger source on its own --- at least one additional non-colocated detector will be required. Possible network configurations were discussed in Section~\ref{ETconfig}. With one additional $10$km detector at the location of LIGO Hanford, the ET network will be able to determine the luminosity distance of a source to an accuracy of $\sim 40\%$. Adding a third $10$km detector at the site of LIGO Livingston or upgrading the detectors to ETs improves this modestly to $\sim 30\%$~\cite{Sesana:2009ET,Gair:2009ET}. There will also be an additional distance error due to weak lensing of the signal, but this will be considerably smaller ($\sim10\%$, see, for example,~\cite{shapiro09}) than the intrinsic error from the gravitational wave observations. If we assume that the luminosity distance is converted into a redshift using the concordance cosmology at that time, the redshift error will be comparable to the distance error. Thus, an ET network should be able to say with confidence if an event is indeed occurring between two $\sim100M_{\odot}$ black holes at high redshift, $z\sim 5$.

\subsection{MBHs in dwarf galaxies}
There are two simple arguments that lead us to believe that $\sim 10^2$--$10^3\,M_\odot$ black holes might inhabit the nuclei of dwarf galaxies today.   Firstly, the mass of SMBHs detected in neighboring galaxies scales with the bulge mass --- or stellar velocity dispersion ($\msigma$) --- of their host galaxy \cite{Magorrian1998,Gebhardt2000,Ferrarese2000,Haring2004}. The lowest-mass galaxies currently known have velocity dispersions $\sigma \sim 10$--$20$ km s$^{-1}$ \cite{Walker2009}. If we extrapolate the $\msigma$ correlation to these $\sigma$ values, we expect the putative IMBHs to have masses in the range of hundreds to thousands of solar masses. 

Secondly, as SMBHs grow from lower-mass seeds, it is natural to expect that a leftover population of progenitor IMBHs should also exist in the present universe.  Indeed, one of the best diagnostics of `seed' formation mechanisms would be to measure the masses of IMBHs in dwarf galaxies. This can be understood in terms of the cosmological bias. The progenitors of massive galaxies have a high probability that the central SMBH is not ``pristine", that is, it has increased its mass by accretion, or it has experienced mergers and dynamical interactions. Any dependence of \mbh\ on the initial seed mass is largely erased.  However, low-mass galaxies undergo a quieter merger history, and as a result, at low masses the BH occupation fraction and the distribution of BH masses still retain
some ``memory'' of the original seed mass distribution. The signature of the efficiency of the formation of SMBH seeds will consequently be stronger in isolated dwarf galaxies \cite{VLN2008}.

One hopes that the next generation of  25-30m optical/IR telescopes operating at their diffraction limit ($\sim 4$ milliarcsec) can provide the first constraints on the presence of IMBHs in dwarf galaxies, but the detection of gravitational waves from a central IMBH in a dwarf galaxy undergoing a merger is possibly a more promising probe. Dwarf galaxies  have a very quiet merger history, hence we do not expect many IMBH-IMBH mergers involving dwarf galaxies at the present epoch, or in the low--redshift universe. The seed black hole mergers discussed in Section~\ref{seeds} probe a separate population of mergers, between the progenitors of galaxies which are more massive today. However, gravitational waves may also be generated in dwarf galaxies by mergers between the central IMBH and stellar remnants in the centre of the dwarf. These are analogous to the globular-cluster IMRI sources described in Section~\ref{GCimri}.

We can derive an estimate of the event rate based on the expected number of dwarf galaxies which can possibly host IMBHs in the interesting mass range. Theoretical models of SMBH formation and evolution, where the seeds of MBHs are Population III remnants \cite{VHM}, can be used to look for the distribution of IMBHs in dwarf galaxies. Using the dynamical model of \cite{vanwass09}, we estimate a number density of IMBHs, $n_{\rm IMBH}\sim 0.02$--$0.1$ Mpc$^{-3}$.

When we calculate the event rate of BH-IMBH mergers in dwarf galaxies, we have to further correct for the fact that only a small fraction of these tiny satellites do indeed form stars \cite{Bovill2009}. Based on \cite{Gnedin2006}, we estimate that a fraction $f_*=0.1-0.2$ of dwarfs in the  $v_{\rm sat}\sim 10-20$ km~s$^{-1}$ range formed stars (which will eventually leave behind stellar mass BHs that can merge with the central IMBH).  
The number density of IMBHs that can be ET sources is therefore $n_{ET}=f_*\,n_{\rm IMBH} \sim 0.001$--$0.06$Mpc$^{-3}$. This number density is about an order of magnitude lower than the number density of globular clusters used to normalise the rates in Section~\ref{GCimri}.

The capture mechanisms that lead to IMRIs in dwarf galaxies are likely to be the same as those that operate in globular clusters. The event rate for the binary-hardening mechanism scales with the stellar density, $n$, as $n^{4/5}$, while the other mechanisms, such as direct capture, should scale approximately with $n$ (this is the same $n$ that enters Eqs.~(\ref{encrate})--(\ref{Thard})). The core stellar densities in nearby dwarf galaxies are typically much lower than in core-collapsed globular clusters, e.g., the estimate for Fornax is $\sim 10^{-1}$pc$^{-3}$~\cite{Mateo:1998} and for Sagittarius is $\sim 10^{-3}$pc$^{-3}$~\cite{Majewski:2005}, compared to $\sim 10^{5.5}$pc$^{-3}$ for globulars~\cite{PryorMeylan:1993}. The IMRI rates for dwarf galaxies are thus likely to be orders of magnitude lower than those for globular clusters. Therefore, although it is not inconceivable that ET will detect events from dwarf galaxies, any events would be serendipitous. Moreover, there are no obvious characteristics which would allow an observer to distinguish between an event in a dwarf galaxy from one in a globular cluster based on the GW signature alone. Nonetheless, it might be possible to make qualitative statements about dwarf galaxy IMBH populations. For instance, if ET does not detect any mergers between seed black holes at high redshift, of the type described in Section~\ref{seeds}, it is very likely that BH seeds were
heavy and not light. This would suggest dwarf galaxies would not contain light leftover BH seeds, and consequently that all of the observed IMRIs
are occurring in globular clusters. Similarly, if seed mergers are detected but the rate of IMRIs is low or zero, it might suggest that IMBH formation
in globular clusters is inefficient and any observed IMRIs are in dwarf galaxies. More refined modelling and detailed calculations are needed to
understand/prove the robustness of these expectations, especially in view of the small number of seed black hole merger events and dwarf
galaxy IMRIs that are predicted. In summary, while the dwarf galaxy channel should not be ignored completely, it is very unlikely to be a significant contributor to ET events or science.

%The central black holes in dwarf galaxy may have masses in the IMBH range.  

\section{Speculative sources \label{spec}}
In this section we discuss some more speculative sources that might be observed by a future low-frequency ground-based interferometer such as ET.  We examine first
the possibility of observing orbiting or rotating white dwarfs
near the high end of allowed masses, then discuss how the observation of 
eccentric compact binaries could illuminate their dynamical origin.

\subsection{Orbiting white dwarfs}

A gravitationally bound object of average density ${\bar\rho}$
has a maximum orbital, rotational, or acoustic frequency
$f_{\rm max}\propto (G{\bar\rho})^{1/2}$.  For neutron
stars this maximum is $\sim 10^3$~Hz.  White dwarfs are much
more extended objects, but near their maximum masses their
densities are sufficient to reach $f_{\rm max}\sim 1$~Hz.
For example, from the classic work \cite{HamadaSalpeter:1961},
a magnesium white dwarf with maximum mass $M_{\rm max}=
1.363~M_\odot$ has a radius $R=2.57\times 10^{-3}~R_\odot=
1.79\times 10^8$~cm and therefore $(G{\bar\rho})^{1/2}=2.7$~Hz.
In the few-Hz range, therefore, one will potentially see
gravitational waves from the most massive white dwarfs.

If we consider specifically such a white dwarf in a binary
orbit with a neutron star, black hole, or another white
dwarf, then the orbital frequency at the point of tidal
disruption of the dwarf depends weakly on the mass of the
companion.  For example, suppose that the equilibrium mass
and radius of the white dwarf are respectively $M_{\rm WD}$ 
and $R_{\rm WD}$, and that the companion is a compact object
of mass $M_{\rm comp}$.  When the orbital separation $a$ is
$a\sim 2R_{\rm WD}(M_{\rm comp}/M_{\rm WD})^{1/3}$, tidal 
stripping begins \cite[and others]{WigginsLai:2000}.  The orbital 
frequency at this point is
\begin{equation}
\omega=\sqrt{G(M_{\rm comp}+M_{\rm WD})/a^3}\sim
0.7(1+M_{\rm WD}/M_{\rm comp})^{1/2}(G{\bar\rho})^{1/2}\; .
\end{equation}
The gravitational wave frequency is $f_{\rm GW}=2f_{\rm orb}=
\omega/\pi$, implying a maximum frequency of $\sim 1$~Hz
for comparable-mass objects such as a neutron star and a
heavy white dwarf, and a maximum that is $\sim$70\% of this
if the companion is a much more massive object such as an IMBH.

We have relatively few candidates for massive white dwarfs,
hence although there is a significant literature related to
lower-frequency radiation from white dwarf binaries
(e.g., \cite{Farmer:2003} and many subsequent papers)
their numbers are difficult to estimate (see
\cite{VennesKawka:2008} for a recent discussion).   Models of
the mass distribution suggest that perhaps $\sim 0.1-1$\% of
white dwarfs have masses near $M_{\rm WD}=M_{\rm max}$
(e.g., see figure 10 of \cite{Catalan:2008}).  Our requirement
that both white dwarfs have masses near the maximum means
that the mass ratio is greater than 2/3, and thus there will
be a merger instead of stable mass transfer
(see \cite{Marsh:2004}).  If we estimate that
that there are $2.5\times 10^8$ double white dwarf systems in
a galaxy like the Milky Way \cite{Nelemans:2001popsynth}, and that 
$\sim 50$\% of the massive ones
have semimajor axes that allow merger by gravitational radiation
within $10^{10}$~yr (corresponding to the $\sim 48$\% merger
fraction from \cite{Nelemans:2001popsynth}), 
then we expect massive white-dwarf binaries to merge
at a rate per galaxy of $\sim (0.001-0.01)^2\times 0.5\times 2.5\times 
10^8/10^{10}\ {\rm yr}^{-1} \sim
2\times 10^{-9}-10^{-6}$~yr$^{-1}$.  At the high end this is similar
to the low end of NS-NS merger rate estimates~\cite{2007PhR...442...75K}.
If the ET is sensitive to such mergers out to $\sim 200$~Mpc, which may be
optimistic given their low GW frequencies, one event per few years could be
detected.  Detection of these events would
indicate rather precisely the maximum average density of white
dwarfs, and would thus be a mechanism for establishing their
mass-radius relation near the maximum mass.

\subsection{Rotating hypermassive white dwarfs}

Another possibility, suggested to us by \cite{Ott:private}, is that two white dwarfs with more typical
masses $M_{\rm WD}<1~M_\odot$ might merge in a binary and
produce a hypermassive white dwarf that spins rapidly enough
that it is deformed into an ellipsoid.  This is a promising
candidate to explain some fraction of Type~Ia 
supernovae~\cite{2009ApJ...699.2026R}.

To evaluate this prospect we note that if a
Newtonian perfect fluid (a good model for a white dwarf) 
rotates uniformly then above a certain critical angular 
momentum, $L_{\rm crit}$, for a given mass, $M$, the equilibrium configuration
splits off from the axisymmetric Maclaurin spheroids (which
emit no gravitational radiation) to the Jacobi ellipsoids.
If the three axes of the ellipsoids are $a_3\leq a_2\leq a_1$,
then, according to \cite[section 39]{Chandrasekhar:1969}, the
critical angular momentum is
\begin{equation}
L_{\rm crit}\approx 0.3(GM^3{\bar a})^{1/2}
\end{equation}
where ${\bar a}\equiv (a_1 a_2 a_3)^{1/3}$.  If two white
dwarfs both of mass $M/2$ and radius $R$ spiral slowly together, 
then their angular momentum at the point of contact is
$L=\mu\sqrt{2GMR}=\sqrt{2}/4(GM^3R)^{1/2}=0.35(GM^3R)^{1/2}$.
Since the equilibrium radius of the hypermassive object is
smaller than the radii of the original white dwarfs, the
angular momentum is sufficient to produce an ellipsoidal
figure.  Again from \cite[section 39]{Chandrasekhar:1969}, the
angular velocity of this configuration will be 
$\Omega\approx (G{\bar\rho})^{1/2}$ and hence the dominant
gravitational wave frequency will be $f_{\rm GW}\approx
(G{\bar\rho})^{1/2}/\pi$.

The amplitude of gravitational waves depends on the ellipticity
$\epsilon\equiv (I_1-I_2)/I_3$, where $I_i$ indicates the moment
of inertia along axis $i$.  Near the
critical angular momentum, slight changes in $L$ produce
large changes in $\epsilon$, and $\epsilon$ of several tenths
is possible.  
Gravitational waves remove rotational energy from the star,
such that
\begin{equation}
{\dot\omega}=-\frac{32}{5}\frac{G}{c^5}\epsilon^2 I_3\omega^5
\end{equation}
where $\omega=\pi f_{\rm GW}$.  As a result, the characteristic
spindown time is
\begin{equation}
T_{\rm spindown}=\omega/|{\dot\omega}|\approx
200~{\rm yr}\left(\frac{0.1}{\epsilon}\right)^2
\left(\frac{10^{49}~{\rm g}~{\rm cm}^2}{I_3}\right)
\left(\frac{f_{\rm GW}}{1~{\rm Hz}}\right)^{-4}\; .
\end{equation}
The sweep rate at 1~Hz is, therefore, $\sim 1~{\rm Hz}/200~{\rm yr}\sim 
10^{-10}~{\rm Hz~s}^{-1}$.  For an integration of
$\sim 10^5$~s the frequency would stay in a single frequency
bin of $\Delta f=1/10^5~{\rm s}=10^{-5}$~Hz. Since the spindown
rate will remain constant for a much longer time a search for a
simple linear drift may make practical integrations over weeks
to months.  This would partially
offset the low expected amplitudes.
For comparison, continuous wave searches in LIGO are routinely done
for spindown times as low as $\sim1000$ years at frequencies of $\sim 100$Hz~\cite{LIGOS5cw}, so
a search for spindown times of $\sim200$ years at $f\sim1$Hz is certainly feasible.

Type~Ia supernovae are estimated to occur once per 1000~years in
galaxies such as the Milky Way~\cite{2009ApJ...699.2026R}, so
even if only 1--10\% of SNe~Ia are binary mergers, the overall
astrophysical rate is competitive with double neutron star
mergers.  Even though the ET sensitivity to gravitational waves
from these binary white dwarf mergers will be much lower than
for double neutron star mergers, the detection of gravitational
waves from any such event may provide a new view on these
important supernovae.

\subsection{Eccentric binaries}

In the sensitivity bands of second-generation gravitational wave
detectors such as Advanced LIGO and Advanced Virgo, most compact
binaries will be very close to circular. (For a proposed scenario
where this may not be true, see \cite{OLeary:2008};  another
possibility includes direct captures of compact objects by IMBHs 
as precursors to eccentric IMRIs in globular clusters, although
this formation mechanism  is uncommon relative to the one
described in Section \ref{GCimri}, which will produce circular
IMRIs).   This is because for moderate to high eccentricities,
gravitational radiation essentially reduces the semimajor axis of
a binary while keeping the pericenter fixed.  Therefore, to have
palpable eccentricity at a given frequency, the pericenter at
formation or at the last dynamical interaction must be inside  the
radius of a circular orbit at that frequency.  For example,  a
binary of two $1.4~M_\odot$ neutron stars must have a pericenter
less than 700~km to be significantly eccentric at a gravitational
wave frequency $f_{\rm GW}=2f_{\rm orb}=10$~Hz.  This is highly
improbable for a field binary, and is even difficult to arrange
for binary-single scattering in dense stellar environments.

Somewhat higher eccentricities can be obtained via the Kozai
secular resonance \cite{Kozai:1962}.  As explored in the context
of black holes by \cite{MillerHamilton:2002b,Wen:2003}, a
binary-binary interaction can result in a stable hierarchical
triple in some tens of percent of encounters.  If the inner binary
and the outer tertiary have orbital planes that are inclined
significantly with respect to each other, then over many orbital
periods the inclination and eccentricity of the inner binary
change periodically, leading at points in the cycle to very small
pericenters and thus potentially observable eccentricity after the
gravitational-wave  driven inspiral.  The eccentricity at 40~Hz is
almost always very small (below 0.1), but at 10~Hz there are a few
orientations in which the eccentricity can be a few tenths
\cite{Wen:2003}.  At still lower frequencies the eccentricity will
be yet higher, because for low eccentricities, $e$, $e\propto
f^{-19/18}$. 

The preceding discussion implies that detector sensitivity at low frequencies will
be important to determine the origin of compact binaries.
In-situ formation from a massive main-sequence binary is still highly
unlikely to produce detectable eccentricities: in order to have eccentricity at
1~Hz, the pericenter distance would have to be $\lesssim 3000$~km 
immediately after the second supernova.
In contrast, dynamical effects such as the Kozai resonance
are expected to produce eccentric orbits at a few Hz.  As a result,
observation of a few BH-BH or BH-NS inspirals at a few Hz
will illuminate their formation processes in a way that is
not as easy at higher frequencies.  We note, however, that simulations
such as those in \cite{OLeary:2006} suggest that of the few per year
to few tens per year of black hole mergers in globulars that are
expected to be seen with Advanced LIGO, less than 10\% are initiated
by the Kozai process.  The greater reach of the Einstein telescope
will enhance the total numbers, but binaries with palpable eccentricity
in the ET band are still expected to be a minority.

%Other signals with EM counterparts?  NS-NS inspirals could be observed for ~1 day, meaning that even a single ET would have some sky location resolution and, possibly, could give early warning for electromagnetic merger observations...  However, more massive sources that we are tasked with discussing will spend little time in band.

\section{Scientific impact of ET observations \label{science}}
ET detections of any of the systems described in this paper will yield important science products, which we now discuss.
 
\subsection{Astrophysics}
The very existence of BHs in the $100$--$1000M_{\odot}$ range is uncertain, so a single robust detection of an IMBH by ET will be of huge significance. If ET detects any seed black hole mergers at high redshift, it will be strong evidence that black hole seeds were {\it light}, which will help discriminate between light and heavy seed scenarios for the growth of structure in the Universe. Observations of mergers between more massive black holes with LISA do not have the same discriminating power, as they cannot distinguish between $\sim10^5M_{\odot}$ MBHs that formed through direct collapse or the collapse of a massive Pop III star and those that formed through a sequence of mergers~\cite{SVH:2007,Gair:2009ET}. A significant number of ET detections of seed black hole mergers may provide constraints on the mass distribution of black hole seeds, and their early accretion history.

Detection of a significant number of IMRIs with ET will indicate that IMBHs form readily in globular clusters (since the rate of IMRIs in dwarf galaxies is so low). The characteristics of the IMRI events will provide constraints on the astrophysics of dense stellar environments, and on the efficiency of capture processes operating within them.

If ET detects white dwarfs undergoing tidal disruption, it will provide important constraints on the physics of degenerate matter, including the maximum density and mass that white dwarfs can reach. Detections of rotating hypermassive white dwarfs would provide information about proposed channels leading to supernovae. Finally, the detection of a significant population of eccentric coalescing binaries will shed light on the efficiency of the processes that drive eccentricity growth in binaries, such as the Kozai mechanism.

\subsection{Fundamental physics}
ET IMRI sources can also be used for testing aspects of relativity theory, in particular verifying that the central object is indeed a black hole as described by the Kerr metric of general relativity. This has been explored extensively in the context of extreme-mass-ratio inspiral events detectable by LISA (see, for example, \cite{Amaro:2007} and references therein).  In the course of an inspiral, the orbit of the smaller object traces out the spacetime geometry of the large body and hence the emitted gravitational waves encode a map of the spacetime structure. One way to characterize this is in terms of the multipole moments of the spacetime. It was demonstrated by Ryan~\cite{Ryan:1995}, for nearly circular and nearly equatorial orbits, that successive multipole moments of an arbitrary spacetime are encoded at different orders in an expansion of the orbital precession frequencies as functions of the orbital frequency. Since these frequencies can be measured from the emitted gravitational waves, a multipole map of the spacetime can in principle be measured. Similar multipole measurements are also possible from observations of ringdown radiation following mergers~\cite{Berti:2006}. For a Kerr black hole, the mass, $M$, and angular momentum, $S$, determine all higher-order mass, $M_l$, and current, $S_l$, multipole moments of the spacetime:
\begin{equation}
M_l + {\rm i}S_l = M({\rm i}S/M)^l.
\label{nohair}
\end{equation}
Measuring just three multipole moments and finding them to be inconsistent with this formula is therefore enough to demonstrate that the central object is not a Kerr black hole.

For IMRIs, it has been shown that Advanced LIGO could measure an $O(1)$ fractional deviation in the mass quadrupole moment, $M_2$, for typical systems~\cite{Brown:2007}. Corresponding results have not yet been computed for ET. However, ET will improve this significantly for two reasons --- (i) the SNR of a source at fixed distance will increase by a factor of $10$ or more; and (ii) ET will observe the sources at lower frequencies. The ability to measure multipole moments improves significantly with the number of gravitational-wave cycles observed. At the leading-order Newtonian approximation, a $1M_{\odot}+100M_{\odot}$ system has $\sim500$ cycles remaining until plunge when the frequency is $10$Hz, but this increases to $\sim1500$ for a frequency of $5$Hz, $\sim4000$ for $3$Hz and $\sim25000$ for a frequency of $1$Hz~\cite{FinnThorne:2000}. ET should thus be able to carry out tests of the Kerr nature of the central object that are significantly better than those possible with Advanced LIGO. Further research is required to quantify the improvement that will be possible, and how this will compare to expected results from LISA EMRI events.

\subsection{Uncertainties}
There are various uncertainties which will affect the scientific impact of ET measurements discussed above. One important consideration is how to distinguish between IMBH events that arise from seed black holes and those that arise from IMBHs formed in globular clusters. Using ET measurements to constrain hierarchical structure formation relies on identification of mergers as seed black hole mergers, but, as we have seen, there may also be IMBH binary mergers in globular clusters. The masses and redshifts of the events may provide a robust discriminator, but more work is needed to understand if this is indeed the case, or whether other characteristic features exist that can be exploited.

The eventual sensitivity that is achieved by ET also has bearing on these results. The speculative sources that were discussed in Section~\ref{spec} rely on ET having sensitivity in the $1$--$10$Hz band, and low-frequency sensitivity should also improve the accuracy of tests of relativity using IMRIs. ET may only have sensitivity down to a frequency of $\sim3$Hz, which will impact all of this science and perhaps eliminate the possibility of detecting gravitational radiation from massive white dwarfs. This must be properly quantified in the future. 

Finally, there are open questions regarding ET data-analysis. The ET data stream will be very source-rich, and so the identification of individual sources of different types in the presence of this confusion will be a challenging problem. For instance, neutron star binary systems will create a confusion background near $1$Hz~\cite{Regimbau:2009}. The data-analysis challenges for ET will inevitably change the SNRs required for detection of individual sources and therefore the rate predictions, and the accuracy with which source parameters can be estimated. However, the rate uncertainties arising from the data analysis will most likely be small compared to the order-of-magnitude uncertainties that are present in the astrophysical rate predictions.

\section{Summary \label{summary}}
We have discussed gravitational waves generated by intermediate-mass black holes as possible sources for the Einstein Telescope. Intermediate-mass black holes may be formed via two alternative channels --- (i) they may be formed in the early Universe if MBH seeds are light (seed IMBH); (ii) they may form in globular clusters via runaway collisions between stars (cluster IMBH). In both cases, there are two distinct types of system that might be sources of gravitational waves for ET --- (a) mergers between binaries containing two IMBHs; (b) mergers of stellar remnants with IMBHs (IMRIs). 

Mergers between seed IMBHs occur following galaxy mergers during the hierarchical assembly of structure. If MBH seeds are light, ET could detect a few to a few tens of seed black hole merger events over three years at redshifts as high as $z\sim8$--$10$. An ET network would, in addition, be able to determine the luminosity distance to these events to an accuracy $\sim30\%$, which is sufficient to say confidently that an event involves {\it intermediate-mass} black holes and is occurring at {\it high redshift}. IMRIs involving seed IMBHs could occur in dwarf galaxies, but the event rate is probably very low, which makes it unlikely that this will be a significant contributor to the ET event rate. If cluster IMBHs form readily, binary IMBHs in globular clusters might be detected by ET at a rate of 
%$\sim$a few$\times 1000$ 
$\sim 500$
per year. Core-collapsed globular clusters are also a more promising host for IMRIs detectable by ET and the IMRI event rate for ET could be as high as a few hundred per year. However, there are significant uncertainties, not least of which is whether IMBHs form at all in the stellar environments of globular clusters.

The improved sensitivity of ET at low frequency may also allow the detection of several speculative sources. High-mass white dwarfs can survive tidal disruption long enough to reach orbital frequencies $f_{\rm max}\sim 1$Hz in binaries. 
%If such systems are detected near the lower frequency cut-off of ET, this will yield constraints on the maximum densities that white dwarfs can reach. 
Hypermassive white dwarfs formed by the mergers of normal white dwarfs in binaries could also be sources for gravitational waves at frequencies around $1$Hz as they will be rapidly rotating and can support relatively significant ellipticities. ET could detect these two types of event at a rate of one per few years, but this number is extremely uncertain.
%Such systems could be associated with a fraction of unusual Type Ia  supernovae. 
Finally, dynamical processes such as the Kozai mechanism can excite %significant 
sufficiently high 
eccentricities in BH-BH and BH-NS binaries, that there would be
% which would lead to
%. While the majority of these systems will have circularized by the time they enter the sensitivity range of Advanced LIGO, they could still have 
significant residual eccentricity when their orbital frequency is in the $1$--$10$Hz range that ET will probe. 
These systems would circularize before reaching orbital frequencies in the Advanced LIGO band. 
%ET detections of significant numbers of eccentric binaries at low-frequency will be an indicator of the efficiency of dynamical formation mechanisms.
ET might detect several eccentric binaries per year, but this rate depends on the fraction of binaries with residual eccentricity and as yet unknown details of the ET data analysis.

ET detections of any of these sources would have significant impact on our understanding of various astrophysical processes, as well as being useful for fundamental physics.

%Summary of previous analysis.
%Testing GR / Kerrness with IMRIs or ringdowns, e.g., by measuring multipole moment structure (see papers by Hughes; Gair et al.; Berti et al.).  
%Astrophysical benefits of such observations with ET.

%\section*{References}

\section*{Acknowledgments}
JG's work is supported by a Royal Society University Research Fellowship.  IM and MV acknowledge support from NASA ATP Grant NNX07AH22G.  MCM acknowledges NASA ATP grant NNX08AH29G.

\bibliographystyle{amsplain}
\bibliography{GRGreview}

\end{document}